\documentstyle[11pt,aaspp4,epsf]{article}
\received{7 December 1998}
\lefthead{Wandel}
\righthead{Baldwin Effect in AGN}

\begin{document}

%personal definitions

\def\myfig #1#2#3#4{\par
\epsfxsize=#1 cm
\moveright #2cm
\vbox{\epsfbox{#3}}
{\noindent Figure~#4 }\vskip .3cm }

\def\apjmsfig #1#2#3#4
{\placefigure{#3}
\figcaption[#3]{#4 
\label{#3}} }
%---------------------------
\def\lg{\rm log}
\def\bh{black hole~}
\def\ew{equivalent width~}
\def\ad{accretion~disk~}
\def\ads{accretion~disks~}
\def\el {emission-line~}
\def\els {emission-lines~}
\def\bhs{black~holes~}
\def\ar{accretion~rate}
\def\ip{ionization-parameter~} 
\def\ed {Eddington~}
\def\ers{{\rm erg/sec}}
\def\ms{M_{\odot}}
\def\et{{\it et al.}}
\def\sw{\rm Schwartzshild~}
\def\bb{black body~}
\def\be{Baldwin effect~}
\def\cmq{cm$^{-3}$}
\def\kms{km s$^{-1}$}

%----------------------

\title{On the Baldwin Effect in Active Galactic Nuclei:}
\title{ I. The Continuum-Spectrum - Mass Relationship}

\author{A. Wandel\altaffilmark{1}}
\affil{ Racah Institute, The Hebrew University, Jerusalem 91904, 
Israel}
\altaffiltext{1}{On leave at Astronomy Department, University of California,
    Los Angeles, CA 90095-1562}

\begin{abstract}
We suggest that the \be is a result of the spectral dependence
 of the line-driving ionizing continuum on the \bh mass.
We derive a relationship between the mass of the central black hole 
and the broad \el luminosity in active galactic nuclei (AGN).
Assuming the UV spectrum of AGN is emitted  from an optically thick medium
we find an expression for the characteristic energy of the ``UV bump'' 
in terms of the observable luminosity and \el width.
We show empirically and analytically that the bump energy 
is anti-correlated with the black-hole mass and with the continuum luminosity.
Our model reproduces the observed inverse correlation between
equivalent width and continuum luminosity,
yielding an explanation of the Baldwin effect from first principles.
The model gives a good fit to the \be of the CIV line for a mean quasar
EUV spectrum (Zheng \et 1997) and for several model spectra.
The model also predicts a correlation between the strength of the \be
(the slope of the \ew as a function of luminosity) and the ionization
potential, consistent with recent data.

\end{abstract}


\section{INTRODUCTION}

Baldwin (1977) showed that the \ew of the CIV \el decreases with increasing 
continuum luminosity, a correlation known as the 'Baldwin effect'. 
This result has been confirmed for a large set of AGN and a wide luminosity
range (Kinney, Rivolo and Koratkar 1990).
A similar
relation has been found also for other broad \el s. 
Several models have been proposed to explain the \be ,
such as a decrease of the ionization parameter or of the covering factor
with luminosity (Wu, Boggess and Gull 1983) or the \ad inclination 
(Netzer 1985). Nevertheless, the origin of the \be
is still not understood (for a recent review see Shields and Osmer, 1998).
 
The widest accepted explanation for the energy source of quasars and Active
Galactic Nuclei (AGN) is the accretion of matter onto a supermassive \bh 
in the center of the host galaxy.
The central mass implied by a variety of observations
(see Wandel 1998) is of the order of $10^6-10^9 \ms $.
The angular momentum directs
the accreted material to form a disk around the central \bh~
and the properties of such an accretion disk may be calculated under a few
basic assumptions (e.g. Sakura and Sunyaev, 1973). In particular one can
calculate the luminosity and spectral distribution of the radiation from
the accretion disk, given the \bh~ mass and the \ar .

The thermal emission from the \ad around a central mass of the size
implied for AGN peaks in the UV, and ionizes the broad \el ~gas. 
The precise shape of the continuum spectrum from an \ad 
depends (among other factors) on the central mass.
Since the continuum radiation ionizes the line-emitting material and
drives the broad AGN \els, the \el strengths depend on the 
continuum spectrum and therefore on the central mass.

 In particular the spectral shape of the ionizing 
radiation can determine the \ew   (the ratio between
the energy flux in the line to the continuum energy flux at the line 
frequency) of the \el s.

We suggest that the \be is a result of the spectral dependence
 of the line-driving ionizing continuum on the \bh mass.
In the next section we present constraints on the \bh~ mass;
section 3 describes the temperature and spectrum of the basic thin \ad model.
Section 4 gives an empirical relation between the model parameters 
($M$ and $\dot M$) and the {\it observed} spectrum and luminosity, 
while section 5 relates these parameters 
to the {\it calculated} spectrum for various \ad models. 
In section 6 we derive the dependence of the line luminosity on the physical
conditions in the gas.
In section in 7 we compose all these elements to produce a relation 
between the \ew and the mass, and in section 8 we show that the model
predictions are indeed consistent with the observed \be for CIV and 
other lines.

\section{Basic relations for accreting Black Holes}

Any compact, accretion powered radiation source must obey several fundamental
relations:

\smallskip
{\bf a. The \ed limit:} 
in order to maintain steady spherical accretion the luminosity must be less than 
the \ed luminosity, 
$$L<L_{Edd}=4\pi GMm_pc/\sigma_T=1.3 10^{46}M_8 ~\ers
$$
or
\begin{equation}
\label {equ:med}
 M_8=0.7 \eta^{-1} L_{46}
\end{equation}

where $ M_8=M/10^8\ms$, $\eta=L/L_{Edd}$ is the \ed ratio, and $L_{46} =L/10^{46}$ \ers .
%For a variable source this upper limit on the \bh~ mass can be mached with a lower limit,
%$M_8<(\Delta t/10^3 {\rm sec}) (R/R_s)^{-1}$

\smallskip
{\bf b. The \bb temperature:} 
if a luminosity $L$ comes from a region of radius R and 
temperature $T$, then  
\begin{equation}
\label {equ:lrt}
L<4\pi R^2\sigma T^4.
\end{equation}

 If a spectral feature at a photon
energy of $E$ is due
to \bb emission from a region of size $R$, then the temperature is given by 
$E\approx 3kT$, and $E\approx (10~eV) (T/10^5K)$,
so that

\begin{equation}
\label {equ:elmr}
E\sim < (10~eV)L_{46}^{1/4}M_8^{-1/2}(R/5 R_s)^{-1/2}.
\end{equation}

 and reversing the relation above
we have an lower limit on the \bh~ mass (for a true \bb spectrum this
becomes an approximate equality):

\begin{equation}
\label {equ:mt}
M_8\sim >2 E_{Ryd}^{-2}{L_{46}}^{1/2} (R/5 R_s)^{-1}
\end{equation}

where $R_s=2GM/c^2\approx 3~10^{13} M_8~$cm is the \sw radius
and $E_{Ryd}$ is the spectral energy in Rydbergs.
(Note that $L_{46}$ is the bolometric luminosity; if the observed
(say optical) luminosity is used, a bolometric correction must be included).

Combining eqs. \ref{equ:med} and \ref{equ:mt} we can eliminate $M$
obtaining (for \bb emission) an estimate of the size:

\begin{equation}
\label {equ:rle}
{R\over R_s}\approx 13 L_{46}^{-1/2} \eta {E_{Ryd}}^{-2}
\end{equation}

\smallskip
{\bf c. Variability:}
the shortest time scale for global variations in the luminosity is the light 
travel time across the \sw radius, hence if the 
luminosity is observed to vary significantly on a time scale $\delta t$, the 
\bh~ mass has to be
\begin{equation}
\label {equ:mdt}
M_8<\frac{\delta t}{10^3 r {\rm sec}},
\end{equation}
where $r$ is the effective radius of emission in units of $R_s$.

However, for continuum changes near the feature  of spectral energy $E$ 
the light travel time is (from eq. \ref{equ:rle})
\begin{equation}
\label {equ:trc}
\delta t_l \approx {R \over c}\approx 1.5 L_{46}^{1/2}  {E_{Ryd}}^{-2} {\rm days},
\end{equation}
and the dynamical time is 
\begin{equation}
\label {equ:trd}
\delta t_d \approx {R\over v} \approx 20 L_{46}^{1/4} \eta^{1/2} {E_{Ryd}}^{-3} {\rm days},
\end{equation}
which gives time scales of the order of the observed  UV variability in 
Seyferts and quasars.
\smallskip

\section{The thin accretion-disk model}
Many authors have tried to fit \ad spectra to the observed AGN continuum, implying the 
\ad parameters ($M, \dot M, \alpha$) and the  \bh~ spin (e.g. \sw or 
Kerr) which give the best fit to 
the observed optical-UV continuum  (e.g. Wandel and Petrosian 1988; Sun and Malkan 
1989; Laor  1990). 
\subsection{The \ad spectrum}

The radiative energy output of the \ad is dominated by the release of gravitational 
potential energy, the rate of which is $GM\dot M/r$. A more accurate 
calculation yields for the flux emitted per unit area
\begin{equation}
\label {equ:lr}
L(R) = {3GM\dot M\over 8\pi R^3} \left [ 1-\left ({R_{in}\over R }\right )^{1/2}\right ]
\end{equation}

where $R_{in}$ is the inner disk radius.
In the outer part of the thin disk the opacity is dominated by true absorption
and the local spectrum is a \bb spectrum. For this part of the disk,
comparing eq. \ref{equ:lr} to the \bb radiation flux gives for the disk surface temperature
\begin{equation}
\label {equ:tr}
T(R) \approx \left (  {3GM\dot M\over 8\pi\sigma R^3} \right )^{1/4} 
%\left ({R\over R_s}\right )^{-3/4}
\approx 6 ~10^5 \left (  {\dot m\over M_8} \right ) ^{1/4} r^{-3/4}~{\rm K}
\end{equation}
\noindent
where $r=R/R_s$ and  
$$\dot m=\dot M/\dot M_{Edd}\approx 2(M/\ms yr^{-1})M_8^{-1}(\epsilon/0.1)^{-1}$$
is the \ar~ in units of the \ed \ar~, $\dot m =1$ at the \ed limit.
When the \ar~ approaches the \ed rate, the intermediate disk region becomes 
dominated by electron-scattering, the spectral function will be of a modified \bb
and the surface temperature is higher than given by eq. \ref{equ:tr}.
At still smaller radii, the pressure in the disk is dominated by radiation
pressure, rather than by 
gas pressure, and the thin disk solution becomes thermally unstable.
In that inner region the thin disk solution is probably not valid, and has to be replaced
by a hot disk solution (e.g. Wandel and Liang 1991).
It turns out that the intermediate modified \bb region is relatively narrow, 
and close to the \bb - electron scattering boundary the spectrum is nearly a \bb one,
so both regimes may be approximated by the \bb solution.

The spectrum of a multi- \bb \ad is given by integrating over the entire
disk, 
\begin{equation}
\label {equ:fn}
L_\nu  \approx \int_{R_t}^{R_{out}} 2\pi R B_\nu [T(R) ] dR
\end{equation}

where $B_\nu(T)$ is the Planck function and
$R_t$ is the transition radius from the intermediate to the inner radiation-pressure
dominated region (for $\dot m<0.02$ the \bb region extends down to the inner edge of the disk
and $R_t=R_{in}$ ).

Since the $B_\nu (T)$ has a sharp peak at $h\nu_{co}\approx 3kT$ and cuts off at higher 
frequencies, the highest frequency of the \bb part of the disk spectrum comes from the 
radius $R_t$, with the highest temperature for which the disk is still optically thick. 
Eq. \ref{equ:fn}
gives a spectrum which depends on the radial extent of the \bb part of the disk.
If that part is extended ($R_{out}/R_t>>1$) the spectrum is almost flat ($\sim \nu^{1/3}$)
and cuts off beyond $h\nu_{co}\approx 3kT(R_t)$ (or $3kT(5R_S)/h$ for $\dot m <0.1$).
If $R_{out}/R_t\sim$ a few,  the spectrum
will be merely a somewhat broadened  Planck spectrum, but if the \ad is extended
over a large radial range there is a significant flat part.
In that case a better approximation than the Planck spectrum is
\begin{equation}
\label {equ:lker}
L_\nu  \approx A \left ({\nu\over \nu_{co}}\right )^{1/3} exp
\left (-{\nu\over \nu_{co}}\right )
\end{equation}
where $A$ and $\nu_{co}$ are the normalization and cutoff frequency
For a Kerr \bh~ Malkan (1990) finds
$\nu_{co} = (2.9\times 10^{15} Hz) \dot M_{0.1}^{1/4}M_8^{-1/2}$,
where $\dot M_{0.1}=\dot M/0.1 \ms yr^{-1}$. This can be written as

\begin{equation}
\label {equ:eco}
h\nu_{co} = (6 ~eV) \dot m^{1/4}M_8^{-1/4}= (20~eV) L_{46}^{1/4}M_8^{-1/2}
\end{equation}
(note that here $L_{46}$ is the {\it observed} luminosity, and we have  included 
a bolometric correction of 10. Also note that for a Kerr \bh the efficiency in the
definition of $\dot m$ is  $\epsilon=0.4$).
The relation between the cutoff frequency and the inner disk temperature
in the two cases of face-on Kerr and the \bb approximation is
$$h\nu_{co} = 0.64kT_{max} = 2.5 kT_{BB}$$ 
(Malkan 1990).

Although the soft X-ray excess suggest a higher thermal temperature in some objects,
In general the characteristic temperature is likely to be related to the UV spectrum; 
While in the UV spectrum is produced by thermal emission from an optically thick gas
(with eventual reproccessing),  the X-rays are more likely to 
be produced by a hotter medium due to processes  such as 
Comptonization, a two-temperature disk (Wandel and Liang 1991) or hot corona 
(Haardt, Maraschi and Ghisellilni 1994; Czerny, Witt and Zycki 1996).

\section{Deriving $M$ and $\dot M$ from the UV spectrum}

We want to find an approximate relation between the \bh~ mass and \ar~ and the 
continuum spectrum.
The \ar~ is determined straightforwardly by the luminosity, via the relation
$L_{ion}=\epsilon f_{ion} \dot M c^2$, where $f_{ion}$ is the fraction of
 the ionizing out of the total luminosity.
To estimate the \bh~ mass we may use the multi-\bb~\ad -spectrum.
As discussed in the previous subsection  the UV bump cutoff frequency is determined by the 
highest surface temperature in the \bb part of the disk.
If the \bb region extends down to a radius $R_t$, and the EUV cutoff energy is $E_{co}$,
then  eq. \ref{equ:mt} gives (we assume the inner region is optically thin and much hotter, 
so its contribution in the frequencies of interest can be neglected)
\begin{equation}
\label {equ:mte}
M_8\approx 2.5 (E_{co} /10 eV )^{-2}L_{46}^{1/2} (R_t/5 R_s)^{-1}.
\end{equation}

In several \ad models the \bb regime extends close to the inner disk edge.
This is the case in the ``$\beta$'' disk model with viscosity proportional to the gas
pressure ($\nu\sim\beta P_{gas}$). For low \ar s ($\dot m<0.1$) this is true also for the
``$\alpha$'' model with viscosity proportional to the total pressure
($\nu\sim\alpha (P_{gas}+P_{rad})$).
In this case, since the disk emissivity peaks at $R\approx 5R_S$ 
(for a non-rotating \bh ) we may
use eq. \ref{equ:elmr} in order to find the cutoff energy. 
\begin{equation}
\label {equ:ecoc}
E\sim >(17~eV)L_{46}^{1/4}M_8^{-1/2}
~~~\sw 
\end{equation}
(where we have introduced a  bolometric correction $L_{opt}=0.1L_{bol}$).

The equivalent result for a Kerr \bh (eq. \ref{equ:eco}) is very similar:
$$E\sim > (20~eV)L_{46}^{1/4}M_8^{-1/2} ~~~{\rm Kerr} .$$

Plotting the cutoff frequency versus the luminosity one can infer
the mass, and vice versa, if the mass can be estimated independently, it is
possible to estimate the cutoff frequency (fig. 1).

\myfig {19} 0 {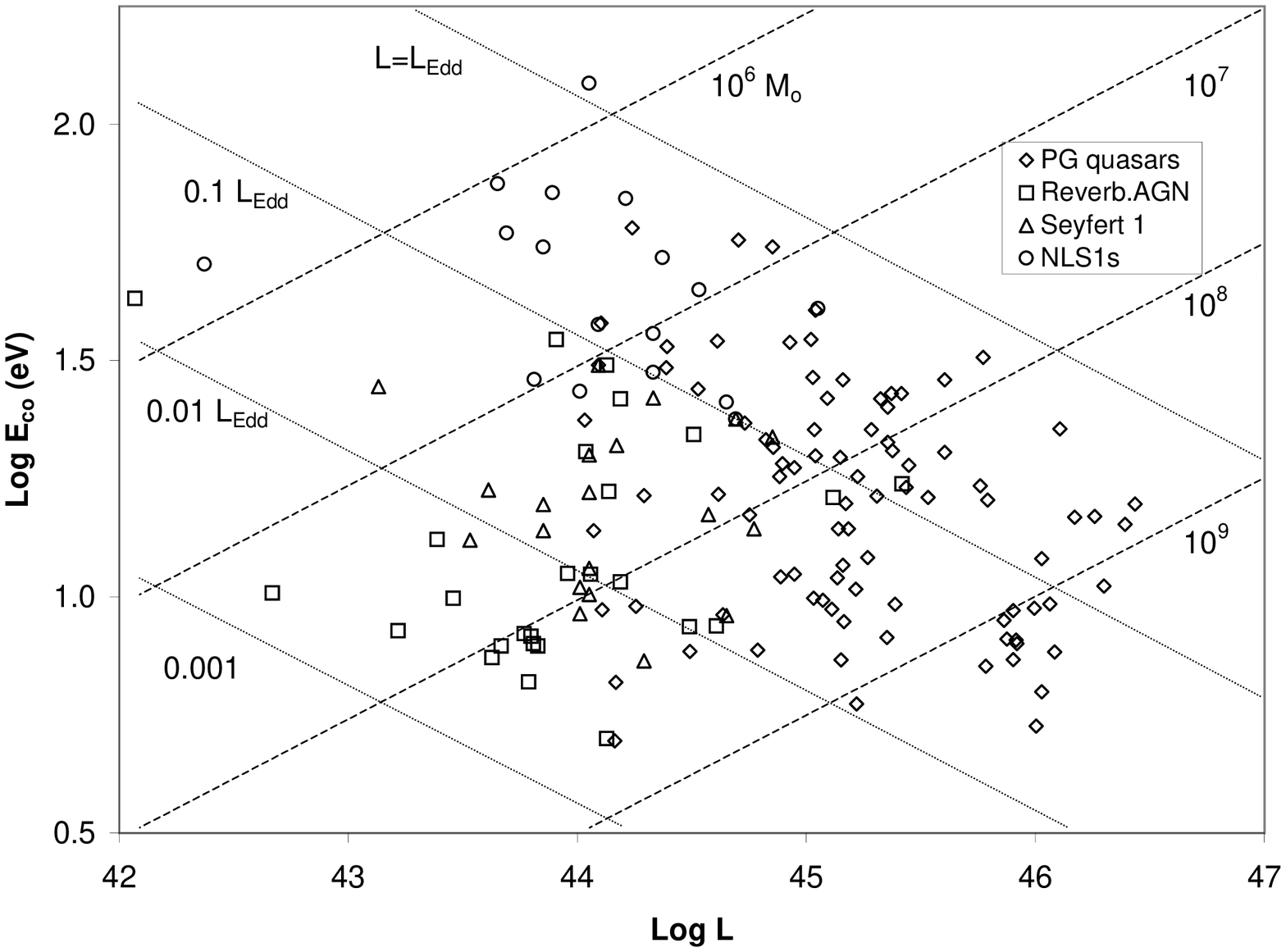}
{1. The cutoff energy (for an \ad spectrum vs. the monochromatic luminosity at
5100\AA. Diagonal dashed lines indicate constant \bh masses, and diagonal 
dotted lines indicate the \ed ratio.
The data shown consists of
 PG quasars (diamonds; from Boroson and Green, 1994), Seyfert 1
galaxies (triangles and squares, the latter indicate AGN with BLR reverberation
data from Wandel, Peterson and Malkan 1999, with luminosities modified for
$H=50$) and Narrow Line Seyfert 1s  
(Boller, Brandt and Fink, 1996) }

It is possible to estimate the \bh mass from the broad emission line profile,
assuming the velocity width is induced by a Keplerian velocity dispersion.
When the broad line region size is estimated by reverberation mapping,
this technique is particularly reliable. Wandel, Peterson and Malkan (1999) have 
used a sample of AGN with reverberation data to calibrate the mass estimate
obtained from photoionization models, finding the relation
\begin{equation}
\label {equ:mph}
M_8\approx 0.4 \left ({L_{46}\over Un_{10}}\right )^{1/2} v_3^2
\end{equation}
where $U$ is the ionization parameter, $n_{10}$ is the density in units
of 10$^{10}$\cmq and $v_3$ is the H$\beta$ FWHM in $10^3$\kms .
Using this calibrated relation makes it possible to estimate \bh mass estimates 
for large samples,
even without reverberation data (Wandel, Malkan and Peterson, in preparation).
Combining equations \ref{equ:mph} and \ref{equ:eco} or \ref{equ:ecoc}   
we have 

\begin{equation}
\label {equ:ecov}
E_{co}\approx (40~eV) (Un_{10})^{-1/4} v_3^{-1}
\end{equation}
Note that
the luminosity dependence cancels out, and the dependence on the unknown
parameters $U$ and $n$ is weak. Using eq. \ref{equ:ecov}
and the average fiducial value of $Un_{10}$=1
 we find the mass and cutoff energy distribution
for a large sample of quasars, Seyfert 1 galaxies and NLS1s (fig 1).
Note that $E_{co}$ is anticorrelated with the mass, and in particular
that different AGN categories group in different regions in the
$L-E_{co}$ plane: quasars have more massive \bhs and low cutoff energies,
Seyfert 1 galaxies have intermediate \bh masses ($10^7-10^8\ms$), and
NLS1s have low \bh masses ($10^6-10^7\ms$) and high cutoff energies.
This is consistent with the large soft X-ray excesses observed for many of
the NLS1s (Boller, Brandt and Fink, 1996).

The diagonal dotted lines in fig. 1 give the $L/L_{Edd}$ ratio - quasars are in the 
0.01-1 range, Seyferts in the 0.001-0.1, and NLS1s are all near \ed,
$L/L_{Edd}\approx 0.1-1$.

A similar analyses for the \sw \ad has been performed by Wandel and Petrosian 
(1988), who calculated for each pair of \ad 
parameters $M$ and $\dot M$  the actual observables, e.g. the UV luminosity and 
spectral index.

\section{Relating $M$ and $L$ to the Ionizing Spectrum}

\subsection{Temperature dependence of the Ionizing Spectrum}

The ionizing power $Q$ is essentially the photon flux above 
the ionization potential $E_{ion}$ of the line in question - 
e.g. 1Ryd (13.6 eV) for Ly$\alpha$ and 48eV for CIV.

We shall now relate the ionizing flux (defined by $Q\propto L_{ion}/<E>$)
to the temperature of the continuum-emitting gas, assuming the emission is
thermal (which is certainly true in the thin \ad model, and supported by the 
presence of the blue
bump feature in  the spectrum of most AGNs.
If  $E_{co}<E_{ion}$  this frequency is on the Wien part, and the flux is sensitively
influenced by the effective disk temperature, $T_{eff}$.
In the parameter range of interest we may therefore assume that the \ew is strongly correlated 
with $T_{eff}$. Quantitatively, the correlation is related to the logarithmic derivative
\begin{equation}
\label {equ:lblt}
{d~ln~B_\nu\over d~ln~T}= {-xe^x\over e^x-1}
\end{equation}
where $x=h\nu /kT$. For $kT<$1 Ryd, which is true in a wide range of cases, as we show below, 
eq. \ref{equ:lblt} gives 
$${d~ln~B_\nu\over d~ln~T}\approx -h\nu/kT,$$ 
namely a strong correlation with T.

In order to relate the UV ionizing spectrum near 1Ryd to the accretion parameters,
it is necessary to determine the effective \bb temperature of the emitting region, 
which we do in the next section.

Once this temperature is determined, we can estimate the ratio of fluxes in any two 
desired energy bands, in particular the flux in the ionizing energy 
band to the energy $E_I$ of the \el in question, which is related the 
\ew , since the line luminosity is related to the ionizing flux.
From eq. \ref{equ:lker} we have 
\begin{equation}
\label {equ:llx}
L_{\nu ion}/L_{\nu I} \approx (E_I/E_{ion})^{1/3}exp[(E_I-E_{ion})/E_o]. 
\end{equation}

\subsection{The Variability-Effective Temperature Relation}

We have already derived a quite general temperature-radius relation for
\bb emission from an accretion flow (eq. \ref{equ:rle}).
We can apply this relation to determine the temperature or radiation peak
energy as a function of the observed luminosity and estimated mass.

In order to determine a characteristic temperature we need either indicate
the size of the emitting region (and use eq. \ref{equ:lrt})
or assume an emission model, as done below for the \ad model.
The size of the emitting region can be bounded by variability analyses.
From eqs. \ref{equ:lrt} and 
\ref{equ:mdt} we have
$$ L<4\pi\sigma c^2(\delta t)^2T^4$$
which gives
\begin{equation}
\label {equ:elt}
E_{eff}> 10 {L_{46}}^{1/4} (t_{day})^{-1/2} eV
\end{equation}
where $t_{day}$ is the variability scale in days.
X-ray data imply a linear relation between luminosity and variability 
time in AGN 
(Barr and Mushotzky, 1986, Ulrich, Maraschi and Urry 1997).
Recently a time scale of the order of $~$30 days has been found in the X-ray 
PDS of NGC 3516 (Edelson and Nandra 1999), but the fastest variability observed
in AGN is of the order of hours.
Assuming a  relation, $t_{day}\sim \tau_L L_{46}$, eq. \ref{equ:elt} gives 
\begin{equation}
\label {equ:el}
E_{eff} > (10~eV)\tau_L^{-1/2} {L_{46}}^{-1/4}.
\end{equation}
For a linear $M-L$ relation $ L_{46}\approx (\eta/0.1) M_8$ (e.g. Wandel and 
Yahil 1985; Wandel, Peterson and Malkan 1999; fig. 1 above) eq. \ref{equ:el} gives
\begin{equation}
\label {equ:emm}
E_{eff}\approx 17\tau_L^{-1/2}(\eta/0.1)^{-1/4}  {M_{8}}^{-1/4} eV,
\end{equation}
similar to the result obtained from \ad models  below.

\subsection{Effective Temperature Models for an \ad}
In the framework of the thin \ad model, 
we consider four models for determining the effective temperature.

{\bf a. The inner disk \bb temperature}

When the disk is nearly \bb up to the inner edge, 
we can approximate the spectrum  by \bb emission near the 
maximum-emissivity radius, which gives
\begin{equation}
\label {equ:ebb}
E_{max}\approx (3~eV) kT_{BB}(5R_s)=10 (\dot m /M_8 )^{1/4} .
\end{equation}
 
As discussed in sec. 3 above, this model applies to the ``$\beta$'' disk and to low
\ar s ($\dot m<0.1$) in the $\alpha$ disk.
A similar expression (eq. \ref{equ:eco}) applies to the Kerr disk.

{\bf b. The inner disk radiation-pressure temperature}

This model applies to high \ar s ($\dot m>0.1$) in the $\alpha$ disk, provided
the inner disk maintains the temperature profile given by the thin-disk solution.
The temperature in the inner, radiation dominated region is given by
(Sakura and Sunyaev 1973)
\begin{equation}
\label {equ:trad}
T_{rad}(r)\approx (2\times10^5~K) (\alpha M_8 )^{-1/4} r^{-3/8} 
[ 1-(6/r)^{1/2}]^{1/4}.
\end{equation}

Substituting the maximal-emissivity radius $r=5$ gives

\begin{equation}
\label {equ:erad}
E_{max}\approx (4~ eV) (\alpha M_8 )^{-1/4} 
\end{equation}
and a higher value for a Kerr \bh .
The gas-radiation pressure boundary -
 taking as $T_{max}$ the temperature at the inner boundary between the \bb and 
the radiation-dominated disk 

{\bf c. The \bb-radiation boundary temperature}

In this case
the effective temperature is taken at the boundary between the intermediate
and inner disk regions,  $r_{mi}\approx 10(\alpha M_8)^{0.1}\dot m^{0.76}$.

The energy of the peak in the spectrum is then derived by substituting $r_{mi}$ 
into eq. \ref{equ:trad}, which gives
\begin{equation}
\label {equ:eradb} 
E_{max}\approx 2 (\alpha\dot m M_8)^{-0.3}~ eV.
\end{equation}

This model is adequate in the case of a high \ar~ - $\alpha$ disk when the inner disk
does not follow the thin disk model but establishes a stable hot solution
(e.g. Wandel and Liang 1991).

 {\bf d. The ``photospheric'' temperature}

In this model we consider as the disk effective temperature 
the temperature at the radius where the disk 
becomes effectively optically thin.
The effective optical depth in the disk is given by

$$\tau_*=h\rho (\kappa_{es}\kappa_a)^{1/2},$$
where $h$ and $\rho$ are the disk vertical scale-height and density, respectively,
and $\kappa_{es}$ and $\kappa_a$ are the electron-scattering and absorption opacities.
Equating $\tau_*$ to unity and eliminating $T$ and $r$ by using the temperature profile 
$T(r)$ gives
\begin{equation}
\label {equ:eradph} 
E_{max}\approx 2 \alpha ^{-0.4}(\dot m M_8)^{-0.26}~ eV.
\end{equation}
Note that the last two models give very similar results.

\section {The Line Emissivity}
The \ew of an \el is actually the ratio between the luminosity in the line and the continuum
luminosity at the \el frequency.
 
The line luminosity is determined by the photoionizing
flux, by the covering factor of the line-emitting gas and by the line emissivity
which is a function of the flux of ionizing photons and the density.
The differential line luminosity emitted from a thin shell can be written as
\begin{equation}
\label {equ:dline}
{d~L_{line}}= L_{ion} {df\over dr} j_{line}(r) dr 
\end{equation}
where $df/dr$ is the differential covering factor, and $j_{line}(Q, n)$, 
the line relative surface emissivity (or reproccessing efficiency), 
is a function of the flux of ionizing photons 
$(Q = L_{ion}/4\pi r^2 c<E>)$, the density ($n$).
$<E>$ is the weighted mean energy per ionizing photon.
The line luminosity is then found by integrating eq. \ref{equ:dline} 
over radius

\begin{equation}
\label {equ:lline}
L_{line}= L_{ion} \int {df\over dr} j_{line}\left [ Q(r),n(r)\right ] dr  
\end{equation}

In order to relate the line luminosity to the ionizing spectrum, we need
first to express the line luminosity in terms of the ionizing luminosity
and spectrum.
The emissivity of several lines has been charted for large portion of the
$Q-n$ (flux and density) plane using the CLOUDY photoionization code in a slab geometry 
(Baldwin \et 1994, Korista \et 1995). These authors find that some lines,
in particular CIV$\lambda 1549$ and
OVI $\lambda 1094$ have a particularly simple emissivity structure, with 
the emissivity peaking strongly (and being approximately constant)
 along the diagonal $Q/n=$const., that is,
for the ionization parameter (defined by $U=Q/n$) being constant, $U\approx U_o$.

Approximating
$$j(Q,n)\approx j_o \delta (U-Uo)$$
(the locally maximally emitting model)
the line luminosity can then be written as

\begin{equation}
\label {equ:j}
L_{line}\approx L_{ion} f_{eff} j_0,
\end{equation}
 $f_{eff}$ is the effective covering factor, given approximately by
integrating $df/dr$ convolved with $j(r)$ over the emitting layer,
and the effective radius (e.g. for calculating $f_{eff}$
is given by the condition $U(r_{eff})=U_0$, or
\begin{equation}
\label {equ:reff}
r_{eff}\times [n(r_{eff})]^{1/2}\approx \left ({L_{ion}\over 4\pi U_0 <E>c}\right )^{1/2}.
\end{equation}

\section {Relating the \ew to $M$ and $L$}

From the definition of the \ew we have
$EW_I\approx L_I/L_c$,
where I denotes the line in question and $L_c$ is the continuum flux at the 
frequency corresponding to that line.
In section 5 we have seen that the thermal UV spectrum from the \ad has a 
strong dependence on the effective surface temperature of the disk, and that 
for a variety of models the functional dependence of the effective photon 
energy on the \bh mass is approximately 

\begin{equation}
\label {equ:emad}
E_{eff}(M)\approx (3-6~eV) M_8 ^{-1/4} 
\end{equation}
where the range in the coefficient accounts for the uncertainty in the 
parameters $\dot m$ and $\alpha$ (for the \ad models) or  $\eta$
and $\tau_L$
(for the variability-temperature relation). 

\subsection{The Temperature-Luminosity Relation}

A more direct relation can be obtained with the luminosity
(which is what we actually need in the context of the \be ).

The variability-temperature relation gives 
$E_{eff}\approx (10 ~eV)(\tau_L/1 day)^{-1/2} L_{46}^{-1/4}$,
For the \ad models we express $M$ and $\dot m$ in terms of $L$.
Since, from the definition of $\dot m$ we have 
$\dot m M_8=1.3\times 10^{46}L_{bol,46}\ers\approx (10-20)L_{opt,46}$,
so that   $$(\dot m M_8)^{1/4}\approx 2 L_{46}^{1/4}$$

The inner \bb model (eq. \ref{equ:ebb}) gives
$$E_{eff}\approx (4~eV)(\dot m/0.1)^{1/2}(\alpha/0.1)^{-1/4} L_{46}^{-1/4}$$
the radiation-pressure model (eq. \ref{equ:erad}) -
$$E_{eff}\approx (8~eV)(\dot m/0.1)^{1/4}(\alpha/0.1)^{-1/4} L_{46}^{-1/4}$$
two last \ad models (eqs. \ref{equ:eradb}, \ref{equ:eradph}) 
can be combined into a single expression
$$E_{eff}\approx (2 ~eV)\alpha^{-1/3}\dot m^{-1/4} M_8^{-1/4}$$
which gives
$$E_{eff}\approx (3~eV) (\alpha/0.1)^{-1/3}L_{46}^{-1/4}.$$

Note that the \ad models are not including Comptonization by a hot
gas, which is likely to increase $E_{eff}$, so they are probably 
underestimating $E_{eff}$.

Combining the variability and \ad estimates we can  write

\begin{equation}
\label {equ:ecomb}
E_{eff}(L)\approx E_c L_{46} ^{-1/4} 
\end{equation}
with $E_c\sim (10-20)$eV,
allowing for a  range in the
parameters $\alpha$, $\tau_L$ and $\dot m$.
This theoretical result is consistent with the empirical distribution 
of $E_{co}$ independently found from the BLR data (fig. 1),
and with direct observations of the 
far UV spectrum of
AGN, which show that the SEDs of several AGN observed with IUE and HUT
peaks at 1000\AA (Zheng \et 1997, 
Kriss \et 1999).

\subsection{Power Law Continuum}

The observed SEDs  suggest that the decrease in the far UV is a power
law ($\sim \nu^{(-(1.5-2)}$ rather than exponential.
Assuming a UV spectrum of the form

$L_\nu\sim  \nu^{-1}$ for $h\nu < E_b$

$L_\nu\sim  \nu^{-\alpha}$ for $h\nu > E_b$

and $E_I<E_b<E_{ion}$
gives

$$
EW_I\propto L_{ion}/\nu L_\nu(E_I)\approx ( E_b/E_{ion})^{\alpha -1}
$$
because  the line luminosity is proportional to the ionizing luminosity
$L_{ion}\approx \nu L_\nu (E_{ion})$.

Taking $E_b=E_{eff}$ we then have
\begin{equation}\label {equ:ewlx}
EW_I\propto L_{ion}/L_I\approx 
 (E_{ion}/E_b)^{-1}\approx (E_c/E_{ion})^{-1}L^{-(\alpha-1)/4}.
\end{equation}
For $\alpha\sim 1.5-2$ (Zheng \et 1997) this gives 
$EW\propto L^{-\gamma}$ with
$\gamma \sim 0.13-0.25$.

While for a pure power law, this would imply the same slope for all lines,
if the spectrum is curving to steeper slopes at higher photon energies
(as could be the case for a partially Comptonized Wien continuum),
lines with higher ionization potential (e.g. OVI) would have a steeper slope
and vice versa - also consitent with the data.

\subsection{Wien Continuum}
From eq. \ref{equ:lker}

\begin{equation}
\label {equ:ewlx}
EW_I=L_{\nu ion}/L_{\nu I} j_0 f_{eff}\propto (E_I/E_{ion})^{1/3}
exp[(E_I-E_{ion})/E_{eff}] 
%\approx (E_I/E_{ion})^{E_{ion}/kT_{eff}}
\end{equation}
%where $E_{eff}=3kT_{eff}$.

Combining these two equations we have
\begin{equation}
\label {equ:ewmx}
EW_I \propto exp \left ({E_I-E_{ion }\over E_c L_{46} ^{-1/4}}\right ) 
\end{equation}

Subject to these uncertainties, we can try to estimate the \be quantitatively.

For the CIV $\lambda$1546 line 
$(E_I-E_{ion})=(8-48)~eV=-40~eV$, which  gives

\begin{equation}../aasn/
\label {equ:ewc4l}
 EW_{CIV} = EW0_{CIV} exp (-40~eV E_c^{-1}  L_{46} ^{1/4} ).
\end{equation}
The coefficient can be determined by normalizing the model to the low
luminosity en to the data, where the \ew seems to be independent of the
luminosity.
Although the data has a large dispersion, part of it is probably due
to intrinsic variability. Plotting the time average \ew vs continuum
luminosity gives a significantly smaller dispersion 
(Kinney, Rivolo and Koratkar 1990, fig 4). From these data we get
$EW0_{CIV}=120~\AA$.

The slope of the \be predicted by eq. \ref{equ:ewc4l} is

\begin{equation}
\label {equ:dewc4dl}
{ dln (EW_{CIV})\over dln L_{46}} =-10 E_c^{-1} L_{46}^{1/4}
\end{equation}

For $E_c=20$eV this gives a slope of $\sim 0.5 L_{46}^{1/4}$.

Fig. 2 shows the data referenced above, 
along with the model prediction (eq. \ref{equ:ewc4l}) as a dashed curve.
We note that the decrease 
in the \ew predicted by eq. \ref{equ:ewc4l} is too large for 
luminous quasars.

\myfig {19} 0 {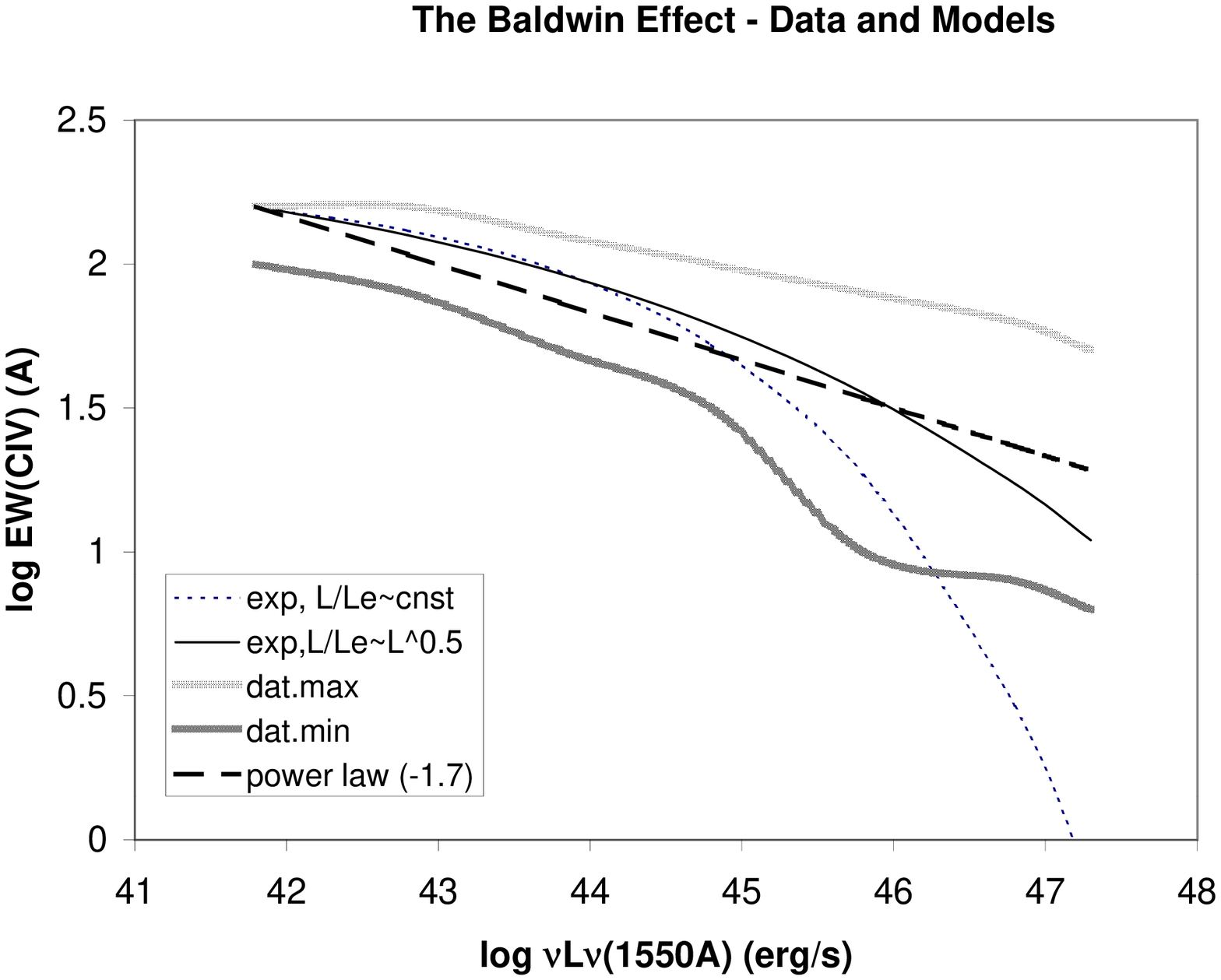}
{2. The \be (\ew of CIV vs. the monochromatic continuum luminosity at
1550\AA. The fuzzy thick curves indicate the dispersion of the data,
taken from Kinney, Rivolo and Koratkar (1990). The smooth curves indicate 
the prediction of the models derived in this work: heavy dashed - 
a power law ionizing spectrum, $L_\nu\sim \nu^{-1.7}$, dotted - with constant 
$L/M$ (eq. \ref{equ:ewc4l} with $E_c=20~eV$) , solid with $L/M\propto L^{1/2}$ (eq. \ref{equ:new}).}

This is a result of our assumption that $L/M\sim constant$.
Actually, as discussed in section 4
the $L/M$ ratio appears to be  correlated with luminosity
(see fig. 1), with the exception of NLS1. 
A similar result is obtained by Wandel and 
Petrosian (1988) and (although on a small range in luminosity)
by Wandel, Peterson and Malkan (1999, fig. 4). 

From fig. 1 we adopt an approximate relation
$$L/L_{Edd}\approx 0.1 L_{46}^{1/2}.$$
Combining this with eqs. \ref{equ:ecoc} yields
$$ E \approx (20~eV) L_{46}^{-1/8}$$
which gives for the \ew
\begin{equation}
\label {equ:new}
 EW_{CIV} = EW0_{CIV} exp (-2  L_{46} ^{1/8} ).
\end{equation}
This result is shown by the solid curve in fig. 2,
which is consistent with the observed \be .

For Ly$\alpha$ we have
$E_I-E_{ion}= 9.2-13.6=4.4$eV, which gives

\begin{equation}
\label {equ:dewladl}
{ dln (EW_{Ly\alpha})\over dln L_{46}} =-1.1 E_c^{-1} L_{46}^{1/4}.
\end{equation}
which is consistent with the weaker \be found for the Ly$\alpha$ line
(Kinney, Rivolo and Koratkar 1990).

\section{Discussion}

\subsection{Comparison with the Data: Luminosity and Energy  dependence}

Our derivation predicts a weak luminosity dependence of the \be .
While this dependence  (eq. \ref{equ:dewc4dl}) on luminosity, 
indicating a slow increase of the \be (or a decrease of the slope
of the line luminosity vs. continuum luminosity) for very luminous quasars
is not obvious from the data, it cannot be ruled out either, because of the
large scatter. 
When the scatter is reduced by averaging multiple observations of the same object,
(e.g. fig. 4a in Kinney, Rivolo and Koratkar (1990)  the resultant \be seems to
suggest an increase the slope of EW plotted vs. L at high luminosities.

We have seen however, that the increase in the \be predicted by eq. 
\ref{equ:dewc4dl}
 is too strong at the highest luminosities.
When the  increase of the \ed ratio with luminosity is taken into account,
the model can fit the CIV data  for the
highest energies, for
 $E_c\approx 20 eV$. This value may be obtained for a disk extending to the 
last stable orbit or by including Comptonization.

Alternatively, the excess effect
could be compensated by the fact that the parameters $\dot m$ and $\tau_L$ 
may depend on luminosity;
If $\dot m$ increases with $L$ (as indicated in Wandel, Peterson and Malkan 1999)
or the variability timescale increase with $L$ less than linearly, 
$E_c$ will increase with luminosity, compensating for the weak luminosity
increase.
An additional uncertainty is due to the unknown 
dependence of the emissivity (which we assume to be constant)
on $L$, which may be influenced, among other factors, by the covering factor.

\subsection{Comparison with the Data: the \be in other lines}

The \be has been observed also for other lines. 
Our model predict a strong dependence of the \be on the ionization energy.
It is interesting to see whether
this prediction could reproduce observed  strength of the \be in different lines. 
Fig. 4b of Kinney, Rivolo and Koratkar (1990) shows a much weaker effect for 
Ly$\alpha$, and in particular for low luminosity objects
($L>10^46\ers$) the Ly$\alpha$ data are consistent with a constant \ew, which
is predicted by our model (eq. \ref{equ:dewladl}).

By the same token, our model predicts that lines with a higher ionization 
potential like OVI) should have  a stronger \be , which is also observed
(Korista \et 1995).

As the emissivity of CIV and OVI has a similar characteristic
distribution in the $Q-n$ plane, one could expect a similar \be
for the two lines, which is indeed observed.
The lack of that sharp feature in the emissivity of lines such as
Ly$\alpha$ and H$\beta$ (Korista \et 1998a) could explain the
fact that these lines do not show a significant \be or only a weak one.
Such a  specific
functional form can be predicted by assuming an explicit radial distribution
of the density and covering factor.
This the expected functional 
form of the Baldwin effect could provide an observational test of the main suggestion of the paper.

\subsection{Normalization}
We have reproduced the \be from the basic relations between \bh mass,
\ad temperature combined with the line emissivity.
It would be of interest to estimate the normalization for the 
proportionality relationship found, which would enable a quantitative 
prediction of the expected \be . However, the
relation between the emission line luminosity and the
ionizing continuum shape depend on the unknown density profile
of the line emitting gas, and on the unknown differential covering factor,
as discussed in section 5.
In the absence of more detailed data on the radial distribution 
of the line emitting gas, further analyses is limited to qualitative
predictions, for which the
positive correlation with the radiation temperature demonstrated in 
sections 5-7 is sufficient. 
A more explicit dependence predicted by photoionization
models can be obtained using the equations derived in sections 5 and 6,
if the radial distribution (characterized by $n(r)$ and $df(r)/dr$)
are known.

\section{Summary}
We suggest that the \be is a result of the distribution of \bh masses
combined with the thermal nature of the line-driving ionizing continuum.
Using the variability-\bb relation, or alternatively the thin \ad  spectrum,
we derive an analytical estimate for the dependence of the continuum spectrum 
on the \bh mass and
show that the ratio of ionizing continuum to the continuum luminosity
near the lines (related to the \ew ) is slowly decreasing with increasing mass.
We  find that the for a model-independent \bb temperature
and for most variants of the thin \ad model
the \ew decreases with increasing continuum luminosity, 
reproducing the observed \be .
 
\acknowledgments
 
Stimulating discussions with Jack Baldwin during
the meeting "Quasars as Standard Candles in Cosmology" at La Serena, Chile, triggered this work.
I also acknowledge valuable discussions with Gary Ferland, Matt Malkan
and Vah\' e Petrosian, the contribution of an anonymous referee,
and the hospitality of the Astronomy Department at UCLA and 
Stanford University.

\end{document}